\documentclass[runningheads]{llncs}
\usepackage{graphicx}
\usepackage{hyperref}
\usepackage{amsmath}
\usepackage{xcolor}
\usepackage{array}

\usepackage{graphicx}
\usepackage{algorithm}
\usepackage{algorithmic}
\usepackage{booktabs}
\usepackage{multirow}
\usepackage{amsfonts}
\usepackage{amssymb}

\usepackage{nicefrac}
\usepackage{url}

\newcommand{\R}{\mathbb{R}}

\makeatletter
\newcommand{\ssymbol}[1]{$^{\@fnsymbol{#1}}$}
\makeatother

\newlength\savewidth

\begin{document}

\title{AlignTransformer: Hierarchical Alignment of \\ Visual Regions and Disease Tags for \\ Medical Report Generation}
\titlerunning{AlignTransformer}
%

\author{Di You\textsuperscript{1}\thanks{Equal Contributions.}, 
Fenglin Liu\textsuperscript{1}$^\star$, 
Shen Ge\textsuperscript{2},
Xiaoxia Xie\textsuperscript{3},
Jing Zhang\textsuperscript{3},
Xian Wu\textsuperscript{2}\thanks{Corresponding Author.}}
\institute{
\textsuperscript{1}School of ECE, Peking University, China\\
\textsuperscript{2}Tencent Medical AI Lab, China \ \ 
\textsuperscript{3}Harbin Chest Hospital, China\\
\email{\{diyou, fenglinliu98\}@pku.edu.cn; \{shenge, kevinxwu\}@tencent.com} \\
\email{\{xiaoxiaxie\_hrb, jingzhang\_hrb\}@foxmail.com}
}
 
\authorrunning{You et al.}

\maketitle              
\begin{abstract}
Recently, medical report generation, which aims to automatically generate a long and coherent descriptive paragraph of a given medical image, has received growing research interests. Different from the general image captioning tasks, medical report generation is more challenging for data-driven neural models. This is mainly due to 1) the serious data bias: the normal visual regions dominate the dataset over the abnormal visual regions, and 2) the very long sequence. To alleviate above two problems, we propose an AlignTransformer framework, which includes the Align Hierarchical Attention (AHA) and the Multi-Grained Transformer (MGT) modules: 1) AHA module first predicts the disease tags from the input image and then learns the multi-grained visual features by hierarchically aligning the visual regions and disease tags. The acquired disease-grounded visual features can better represent the abnormal regions of the input image, which could alleviate data bias problem; 2) MGT module effectively uses the multi-grained features and Transformer framework to generate the long medical report. The experiments on the public IU-Xray and MIMIC-CXR datasets show that the AlignTransformer can achieve results competitive with state-of-the-art methods on the two datasets. Moreover, the human evaluation conducted by professional radiologists further proves the effectiveness of our approach.


\end{abstract}

\section{Introduction}
\begin{figure}[t]

\centering
\includegraphics[width=1\linewidth]{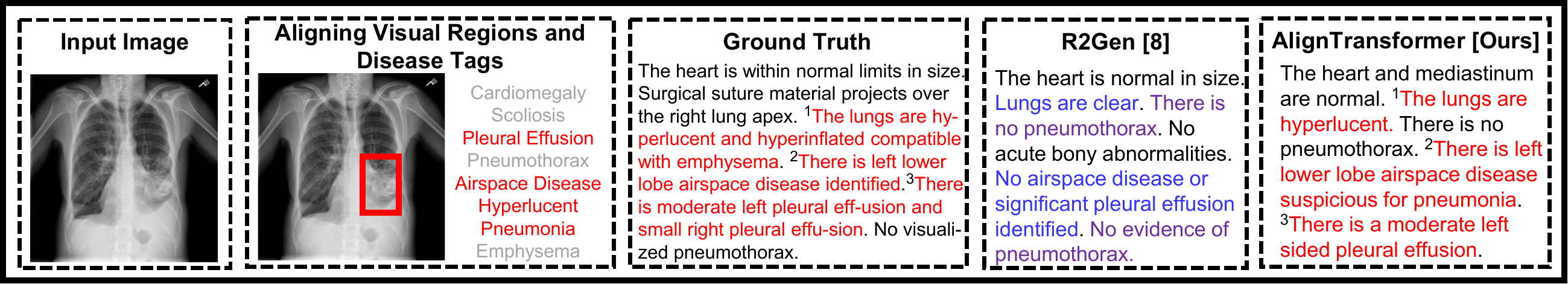}
\vspace{-15pt}
\caption{An example of ground truth report and reports generated by a state-of-the-art approach R2Gen \cite{Chen2020Generating} and our approach. The {\color{red} Red} bounding boxes and {\color{red} Red} colored text indicate the abnormalities in images and reports, respectively. As we can see, in this example, the R2Gen fails to depict the rare but important abnormalities and generates some error sentences ({\color{blue} Blue} colored text) and repeated sentences ({\color{purple} Purple} colored text).}
\vspace{-10pt}
\label{fig:introduction}
\end{figure}

Medical images, e.g., radiology and pathology images, as well as their corresponding reports (see Fig.~\ref{fig:introduction}) are widely-used for diagnosis \cite{Delrue2011Difficulties,goergen2013evidence}.
In clinical practice, writing a medical report can be time-consuming and tedious for experienced radiologists, and error-prone for inexperienced radiologists \cite{Brady2012DiscrepancyAE,fenglin2021Contrastive,fenglin2021PPKED}.
As a result, automatic medical report generation systems can reduce the workload of radiologists by assisting them in clinical decision-making, and are thus in urgent need \cite{Jing2018Automatic,Li2018Hybrid,Jing2019Show,Li2019Knowledge,Xue2018Multimodal,Yuan2019Enrichment,Biswal2020CLARA,Tanveer2020Chest,Zhang2020When,Chen2020Generating,fenglin2021Contrastive,fenglin2021PPKED,fenglin2021CMCL}.

Most existing medical report generation models follow the standard image captioning approaches \cite{Xu2015Show,Vinyals2015Show,You2016Image,anderson2018bottom,fenglin2018simnet,fenglin2019GLIED,fenglin2019MIA,Pan2020XLinear,Cornia2020M2} and employ the encoder-decoder framework, e.g., a CNN-based image encoder followed by a LSTM-based report decoder.
However, directly applying image captioning approaches to medical images has the following problems:
1) \textbf{Data bias}: the normal images dominate the dataset over the abnormal ones \cite{Shin2016Learning}. Furthermore, for each abnormal image, the normal regions dominate the image over the abnormal ones \cite{fenglin2021CMCL}. As shown in Fig.~\ref{fig:introduction}, abnormal regions (Red bounding boxes) only occupy a small part of the entire image.
As a result, existing models tend to generate plausible general reports with no prominent abnormal narratives, failing to depict the rare but important abnormalities \cite{Jing2019Show,Li2018Hybrid,Yuan2019Enrichment}.
2) \textbf{Very long sequence}: Image captioning models are designed to generate one single and short sentence, which only describes the most prominent visual contents,
while medical report generation requires to generate a long paragraph, including multiple structural sentences with each one focusing on a specific medical observation.
For the widely-used LSTM \cite{Hochreiter1997LSTM} in image captioning, it is hard to model such long sequences or paragraphs due to vanishing or exploding gradients \cite{Pascanu2013difficulty}.
Recent medical report generation models \cite{Jing2018Automatic,Jing2019Show,Yuan2019Enrichment} rely on hierarchical LSTM \cite{Krause2017Hierarchical}, which have similar long sequence modeling problems.
As shown in Fig.~\ref{fig:introduction}, even a state-of-the-art model R2Gen \cite{Chen2020Generating} still generates some repeated sentences of normalities.

To alleviate above problems, we propose the AlignTransformer, which includes two major modules: 1) Align Hierarchical Attention (AHA) works as the encoder to extract visual features. To focus on the abnormal regions, AHA first predicts the disease tags from the input image and then aligns these tags with the corresponding visual regions. To fit for both coarse- and fine-grained visual features, AHA introduces a hierarchically alignment model \cite{fenglin2019MIA}.
In detail, the disease tags are firstly used to find the most relevant visual regions, extracting disease-grounded visual features for each disease tag. Then the extracted visual features are further used to find the most relevant disease tags and filter out the irrelevant disease tags for each visual feature.
By conducting the above process at different granularity level, the visual receptive fields gradually concentrate on salient abnormal regions under the guidance of the disease tags.
In this way, we can obtain the features of abnormal regions, which can be used to alleviate data bias; 2) Multi-Grained Transformer (MGT) works as the decoder to generate final reports. MGT adaptively exploit coarse- and fine-grained disease-grounded visual features through a learned gating mechanism.
In addition, MGT introduces the Transformer \cite{Vaswani2017Transformer,Cornia2020M2,Chen2020Generating,fenglin2021PPKED} to generate the reports. Benefiting from the visual features from multiple granularities and the Transformer, as shown in Fig.~\ref{fig:introduction}, our approach are a better fit for long report.

Overall, the contributions of this paper are as follows:
\begin{itemize}
    \item In this paper, we propose the AlignTransformer framework, which aims to alleviate the data bias problem and model the very long sequence for medical report generation.

    \item The experiments on the benchmark IU-Xray \cite{Dina2016IU-Xray} and MIMIC-CXR \cite{Johnson2019MIMIC} datasets prove the effectiveness of our approach, which achieves results competitive with the existing state-of-the-art methods. We also verify the advantage of AlignTransformer from the manual evaluation of radiologists. 

\end{itemize}

\section{Related Works}
The related works are introduced from: 1) Image Captioning and Visual Paragraph Generation and 2) Medical Report Generation.

\smallskip\noindent\textbf{Image Captioning and Visual Paragraph Generation}
Image captioning \cite{chen2015microsoft} aims to understand the given images and generate corresponding descriptive sentences \cite{Xu2015Show,Vinyals2015Show,anderson2018bottom,fenglin2018simnet,fenglin2019GLIED,fenglin2019MIA,Pan2020XLinear,Cornia2020M2}.
However, the sentence generated by AlignTransformer is usually short and describes the most prominent visual contents, which cannot fully convey the rich feature information of the image.
Recently, visual paragraph generation \cite{Krause2017Hierarchical}, which aims to generate long and coherent reports or stories to describe visual contents, has attracted increasing research interests. 
To conduct the task, the Hierarchical LSTM structure (HLSTM) \cite{Krause2017Hierarchical} is widely-used.
However, in paragraph generation for medical images, the correctness of generating abnormalities should be emphasized more than other normalities, while in paragraphs of natural images each sentence has equal importance.
Besides, due to the data bias in the medical domain, the widely-used HLSTM in the visual paragraph generation does not perform very well in medical report generation and is tend to produce normal reports \cite{Xue2018Multimodal,Li2018Hybrid,Yuan2019Enrichment,Jing2019Show,fenglin2021Contrastive,fenglin2021PPKED,fenglin2021CMCL}.

\smallskip\noindent\textbf{Medical Report Generation}
Inspired by the success of deep learning models on image captioning, a lot of encoder-decoder based frameworks have been proposed \cite{Jing2018Automatic,Jing2019Show,Yuan2019Enrichment,Xue2018Multimodal,Li2018Hybrid,Zhang2020When,Li2019Knowledge,Liu2019Clinically,Chen2020Generating,fenglin2021Contrastive,fenglin2021PPKED,fenglin2021CMCL} to automatically generate a fluent medical report.
However, due to the data bias and the very long sequence, some errors occur in the generated reports of the existing methods, like duplicate reports, inexact descriptions, etc \cite{Xue2018Multimodal,Yuan2019Enrichment,Jing2018Automatic}.
To this end, \cite{Jing2019Show,Li2018Hybrid,Liu2019Clinically} and \cite{fenglin2021PPKED,Zhang2020When,Li2019Knowledge} introduced the reinforcement learning and medical knowledge graph, respectively.
\cite{fenglin2021Contrastive} introduced the contrastive attention to compare the abnormal image with the normal images to better capture the abnormalities.
Meanwhile, \cite{fenglin2021PPKED,Chen2020Generating} further introduced the Transformer \cite{Vaswani2017Transformer} to model the very long sequence.
In our work, different from \cite{fenglin2021PPKED,Chen2020Generating}, we conduct the alignments between visual regions and disease tags in \textit{an iterative manner} and propose \textit{a multi-grained transformer} to combine disease-grounded visual features at different depths.

\section{Approach}
We first formulate the problem; Next, we introduce our framework, which includes the Align Hierarchical Attention and Multi-Grained Transformer.

\vspace{-5pt}
\subsection{Problem Formulation}
Given a medical image encoded as visual features $V$ and disease tags $T$, the goal of our framework is to generate a coherent report $R$ that describes the observations in details of both normal and abnormal regions.
In implementations, for the visual features $V$, we follow \cite{Jing2019Show,Li2019Knowledge,Li2018Hybrid,fenglin2021Contrastive,fenglin2021PPKED,fenglin2021CMCL} to adopt the ResNet-50 \cite{he2016deep} pre-trained on ImageNet \cite{Deng2009ImageNet} and fine-tuned on CheXpert dataset \cite{Irvin2019CheXpert} to extract the 2,048 $7 \times 7$ feature maps, which are further projected into 512 $7 \times 7$ feature maps, denoted as $V = \{v_1, v_2, \dots, v_{N_\text{V}}\} \in \R^{N_\text{V} \times d}$ ($N_\text{V}=49, d=512$).
Moreover, we follow \cite{Jing2018Automatic} to further predict the disease tags $T$ of the input image.
Specifically, we further feed the extracted visual features $V$ into a multi-label classification network, which is pre-trained as multi-label classification task on the downstream datasets to generate a distribution over all of the pre-defined tags.
Finally, the embeddings of the $N_\text{T}=10$ most likely tags $T=\{t_1,t_2,\dots,t_{N_\text{T}}\} \in \R^{N_\text{T} \times d}$ are used as disease tags of current input image.
Based on the extracted $V$ and $T$, our framework first adopt the AHA to align the visual regions $V$ and disease tags $T$ in a coarse-to-fine manner, acquiring multi-grained disease-grounded visual features $\hat{V}$. Next, MGT is introduced to generate final report based on the $\hat{V}$.
In brief, our AlignTransformer framework is formulated as:
\begin{equation}
\label{eqn:formula}
\footnotesize
\text{AHA} : \{V,T\} \to \hat{V} ;
\quad
\text{MGT} : \hat{V} \to R .
\end{equation}
Through the above process, our framework gets the ability to alleviate the data bias problem and model the very long sequence.

\vspace{-5pt}
\subsection{Align Hierarchical Attention}
\subsubsection{Basic Module}
To align the visual regions $V$ and the disease tags $T$, we adopt the Multi-Head Attention (MHA) from \cite{Vaswani2017Transformer}, which allows probabilistic many-to-many relations instead of monotonic relations, as in \cite{fenglin2019MIA,Xu2015Show}. 
Thus, the MHA can compute the association weights between different features.
The MHA consists of $n$ parallel heads and each head is defined as a scaled dot-product attention:
\begin{align}
\label{eqn:MHA}
\text{Att}_i(X,Y) &= \text{softmax}\left(X\text{W}_i^\text{Q}(Y\text{W}_i^\text{K})^T\right)Y\text{W}_i^\text{V}
\\
\text{MHA}(X,Y) &= [\text{Att}_1(X,Y); \dots; \text{Att}_n(X,Y)]\text{W}^\text{O}
\end{align}
where $X \in \R^{l_x \times d}$ and $Y \in \R^{l_y \times d}$ denote the Query matrix and the Key/Value matrix, respectively; the divisor $\sqrt{{d}_{n}}$ (${d}_{n}=d/n$) in Eq.~(\ref{eqn:MHA}) is omitted in equations for conciseness, please see \cite{Vaswani2017Transformer} for details; $\text{W}_i^\text{Q}, \text{W}_i^\text{K}, \text{W}_i^\text{V} \in \R^{d \times d_n}$ and $\text{W}^\text{O} \in \R^{d \times d}$ are learnable parameters. $[\cdot;\cdot]$ stands for concatenation operation. 

Conventionally, a Fully-Connected Network (FCN) is followed by the MHA:
\begin{align}
\text{FCN}(X) = \max(0,X\text{W}^\text{f}+\text{b}^\text{f})\text{W}^\text{ff}+\text{b}^\text{ff} 
\end{align}
where $\max(0,*)$ represents the ReLU activation function;
$\text{W}^\text{f} \in \R^{d \times 4d}$ and $\text{W}^\text{ff} \in \R^{4d \times d}$ are learnable matrices; $\text{b}^\text{f}$ and $\text{b}^\text{ff}$ are the bias terms. 
It is worth noting that both the MHA and the FCN are followed by an operation sequence of dropout \cite{srivastava2014dropout}, skip connection \cite{he2016deep} and layer normalization (Norm) \cite{ba2016layernormalization}.

\begin{figure}[t]

\centering
\includegraphics[width=0.9\linewidth]{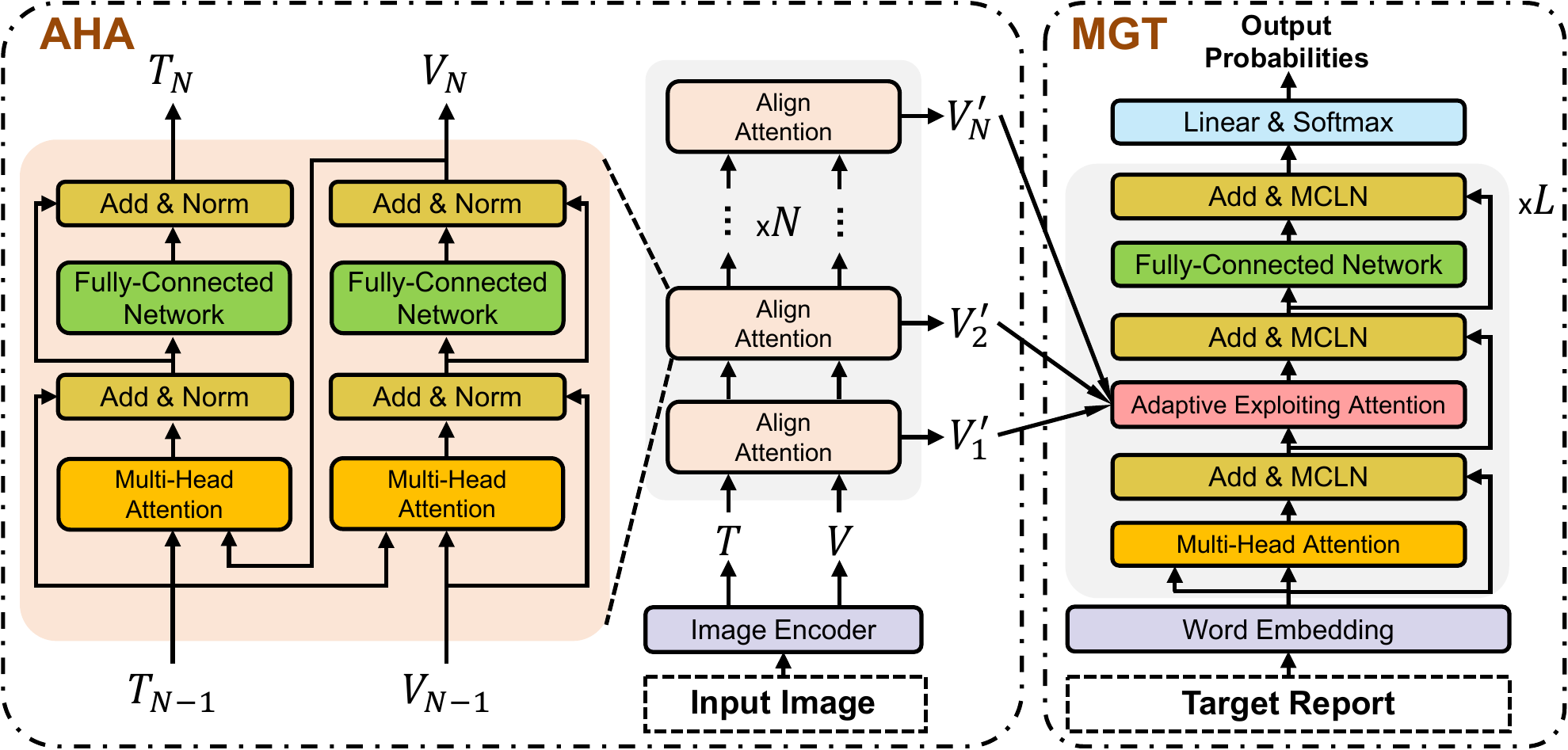}
\vspace{-1pt}
\caption{Illustration of our proposed AlignTransformer framework, which consists of the Align Hierarchical Attention (AHA) and Multi-Grained Transformer (MGT).}
\vspace{-2pt}
\label{fig:framework}
\end{figure}

\smallskip\noindent\textbf{Align Attention}
Since the MHA can compute the association weights between different features, if we apply the MHA between visual regions $V$ and the disease tags $T$, we can learn the correlation and relationship between $V$ and $T$, resulting in the alignment between visual regions and disease tags.
Therefore, based on the MHA, as shown in Fig.~\ref{fig:framework}, the align attention is conducted as:
\begin{align}
\label{eqn:align}
V' = \text{FCN}(\text{MHA}(T, V)) ; \quad T' = \text{FCN}(\text{MHA}(V', T))
\end{align}
In implementation, the disease tags $T$ are first used as the Query to find the most relevant visual regions in $V$, generating the disease related visual features $V'$. 
Then the $V'$ are further used as the Query to find the most relevant disease tags and filter out the irrelevant disease tags in $T$.
As a result, we can obtain alignment between visual regions and disease tags via above processes.
Finally, since $V'$ and $T'$ are aligned, we can add them up to get the disease-grounded visual features $\hat{V}$ that exploits the best of their respective advantages:
\begin{align}
   \hat{V} = \text{LayerNorm}(V' + T')
\end{align}
where the LayerNorm denotes the layer normalization \cite{ba2016layernormalization}.
In particular, since $T$ contains the disease tags, the disease-grounded visual features $\hat{V}$ can be referred as the features of abnormal regions, which are useful in alleviating data bias.

\smallskip\noindent\textbf{Full Model}
To improve the performance and acquire the multi-grained disease-grounded visual features, we propose to perform Align Attention hierarchically, which we call Align Hierarchical Attention.
Specifically, the process in Eq.~(\ref{eqn:align}) that uses the original features is considered as the first round:
\begin{align}
V_1 = \text{FCN}(\text{MHA}(T_0, V_0)) ; \quad T_1 = \text{FCN}(\text{MHA}(V_1, T_0))
\end{align}
where $V_0=V$, $T_0=T$, $V_1$ and $T_1$ denote original visual features, original disease tags, aligned visual features, and aligned disease tags, respectively. By repeating the same process for $N$ times, we obtain the final outputs of the two stacks:
\begin{align}
V_N = \text{FCN}(\text{MHA}(T_{N-1}, V_{N-1})) ;& \quad T_N = \text{FCN}(\text{MHA}(V_{N}, T_{N-1})) \\
\hat{V}_N =  \text{Layer}&\text{Norm}(V_N + T_N) 
\end{align}
In this way, we obtain the multi-grained disease-grounded visual features $\{\hat{V}_i\}=\{\hat{V}_1, \hat{V}_2, \dots, \hat{V}_N\}$.

\vspace{-5pt}
\subsection{Multi-Grained Transformer}
We propose the Multi-Grained Transformer (MGT), which includes $L=3$ decoder layers, to exploit the $\{\hat{V}_i\}$ and generate a proper report $R$.
In implementations, for generating each word $y_t$, given the embedding of current input word $x_t^{(0)} = w_t + e_t$ ($w_t$: word embedding, and $e_t$: fixed position embedding). For the $l^\text{th}$ decoder layer, it takes $x_t^{(l-1)}$ as input to obtain $h_t^{(l)} = \text{MHA}(x_t^{(l-1)}, x_{1:t}^{(l-1)})$.
Then, inspired by \cite{Cornia2020M2}, we introduce the Adaptive Exploiting Attention (AEA) module to adaptively exploit multi-grained disease-grounded visual features:
\begin{align}
\hat{h}_t^{(l)} =& \text{AEA}(h_t^{(l)}, \{\hat{V}_i\}) = \sum_{i=1}^{N}
\lambda_i \odot \text{MHA}(h_t^{(l)}, \hat{V}_i)
\\
\lambda_i &= \sigma\left(\left[h_t^{(l)}; \text{MHA}(h_t^{(l)}, \hat{V}_i)\right]\text{W}_i + b_i\right)
\end{align}
where $\text{W}_i \in \R^{2d \times d}$ is a learnable parameter and $b_i$ is a learnable bias vector. $\odot$, $\sigma$ and $[\cdot;\cdot]$ denote the element-wise multiplication, the sigmoid function and the concatenation, respectively.
The computed $\{\lambda_i\}$ weights the expected importance of different grained disease-grounded visual features for each target word.
Then, we adopt the FCN to acquire the output of $l^\text{th}$ decoder layer: $x_t^{(l)} = \text{FCN}(\hat{h}_t^{(l)})$
It is worth noting that both the MHA, the AEA and the FCN are followed by an operation sequence of dropout \cite{srivastava2014dropout}, skip connection \cite{he2016deep} and memory-driven conditional layer normalization (MCLN) \cite{Chen2020Generating}.

Finally, the output of last decoder layer $x_t^{(L)}$ is used to predict the next word:
$y_{t} \sim p_{t}=\text{softmax}(x_t^{(L)}\text{W}^p + \text{b}^p)$,
where the $\text{W}^p$ and $\text{b}^p$ are the learnable parameters. 
Given the ground truth report $R^*=\{y^*_1,y^*_2,\dots,y^*_{N_\text{R}}\}$ provided by the radiologists, we can train our AlignTransformer framework by minimizing the widely-used cross-entropy loss:
$L_{\text{CE}}(\theta)=-\sum_{i=1}^{N_\text{R}} \log \left(p_{\theta}\left(y_{i}^{*} \mid y_{1: i-1}^{*}\right)\right)$.

\section{Experiments} 
\subsubsection{Datasets, Metrics and Settings}  In our work, we conduct experiments on two public datasets, i.e., a widely-used benchmark IU-Xray \cite{Dina2016IU-Xray} and a recently released large-scale MIMIC-CXR \cite{Johnson2019MIMIC}.
\textbf{IU-Xray} is widely-used to evaluate the performance of medical report generation methods. It contains 7,470 chest X-ray images associated with 3,955 reports. Following previous works \cite{Chen2020Generating,Jing2019Show,Li2019Knowledge,Li2018Hybrid,fenglin2021Contrastive,fenglin2021PPKED,fenglin2021CMCL}, we randomly split the dataset into 70\%-10\%-20\% training-validation-testing splits.
\textbf{MIMIC-CXR} is the recently released largest dataset to date and consists of 377,110 chest X-ray images and 227,835 reports from 64,588 patients. Following \cite{Chen2020Generating,fenglin2021Contrastive,fenglin2021PPKED,fenglin2021CMCL}, we use the official splits to report our results, resulting in 368,960 in the training set, 2,991 in the validation set and 5,159 in the test set.

To test the performance, we adopt the widely-used BLEU \cite{papineni2002bleu}, METEOR \cite{Banerjee2005METEOR} and ROUGE-L \cite{lin2004rouge}, which are reported by the evaluation toolkit \cite{chen2015microsoft}.
Since our approach is based on the Transformer \cite{Vaswani2017Transformer}, we follow the original settings in \cite{Vaswani2017Transformer} and set $d=512, n=8$.
Based on the average performance on the validation set, both the $N$ and $L$ in our Align Hierarchical Attention and Multi-Grained Transformer, respectively, are set to $3$.
We use Adam optimizer \cite{kingma2014adam} with a batch size of 16 and a learning rate of 2e-4 within maximum 100 epochs for parameter optimization.
We also use momentum of 0.8 and weight decay of 0.999.

\begin{table}[t]
\centering
\scriptsize
\caption{Results of our AlignTransformer and existing models on the MIMIC-CXR and IU-Xray datasets. Hit represents the human evaluation results and is measured by the pick-up percentages (\%). The differences between their sum and 100\% represent the percentage of ties.
Higher value denotes better performance in all columns. 
\label{tab:main_result}}
\setlength{\tabcolsep}{2pt}
\begin{tabular}{@{}l l c c c c c c c c@{}}
\toprule 

Datasets & Methods & Year & BLEU-1 & BLEU-2 & BLEU-3 & BLEU-4 & METEOR & ROUGE-L & Hit\\
\midrule [\heavyrulewidth]
\multirow{7}{*}[-2pt]{\begin{tabular}[c]{@{}l@{}} MIMIC \\ -CXR \end{tabular}}
& CNN-RNN \cite{Vinyals2015Show} & 2015 &0.299 &0.184 &0.121 &0.084 &0.124 &0.263 &-  \\
& AdaAtt \cite{lu2017knowing} & 2017 &0.299 &0.185 &0.124 &0.088 &0.118 &0.266 &-\\
& Att2in \cite{rennie2017self} & 2017 &0.325 &0.203 &0.136 &0.096 &0.134 &0.276 &-\\ 
& Up-Down \cite{anderson2018bottom} & 2018 & 0.317 &0.195 &0.130 &0.092 &0.128 &0.267 &- \\
& R2Gen \cite{Chen2020Generating} & 2020 &0.353 &0.218 &0.145 &0.103 &0.142 &0.277 & 29 \\
& PPKED \cite{fenglin2021PPKED} & 2021 & 0.360 & 0.224 & 0.149 & 0.106 & 0.149 & \bf 0.284 &- \\
\cmidrule(l){2-10}
& AlignTransformer & Ours & \bf 0.378 & \bf 0.235 & \bf 0.156 & \bf 0.112 & \bf 0.158 & 0.283 & \bf 54 \\ \midrule

\multirow{7}{*}[-2pt]{IU-Xray} 
& HRGR-Agent \cite{Li2018Hybrid} & 2018 &0.438 & 0.298 & 0.208 & 0.151 & - & 0.322 & -  \\
& CMAS-RL \cite{Jing2019Show} & 2019 & 0.464 &0.301 &0.210 &0.154 & - & 0.362 & - \\
& SentSAT \cite{Zhang2020When} & 2019 & 0.445 & 0.289 & 0.200 & 0.143 & - & 0.359 & -  \\
& SentSAT+KG \cite{Zhang2020When} & 2020 &0.441 &0.291 &0.203 &0.147 & - &0.367 & - \\
& R2Gen \cite{Chen2020Generating} & 2020 &0.470 &0.304 &0.219 &0.165 & 0.187 & 0.371 & 22 \\
& PPKED \cite{fenglin2021PPKED} & 2021 & 0.483 & \bf 0.315 & 0.224 & 0.168 & 0.190 & 0.376 &- \\
\cmidrule(l){2-10}
& AlignTransformer & Ours & \bf 0.484 &  0.313 & \bf 0.225 & \bf 0.173 & \bf 0.204 & \bf 0.379 & \bf 63 \\ 
\bottomrule
\end{tabular}
\end{table}

\smallskip\noindent\textbf{Automatic Evaluation} \ 
Several representative models, including three most recently published state-of-the-art models, i.e., PPKED \cite{fenglin2021PPKED}, R2Gen \cite{Chen2020Generating} and SentSAT + KG \cite{Zhang2020When}, are selected for comparison.
The results on the test of two datasets are shown in Table~\ref{tab:main_result}.
As we can see, our AlignTransformer achieves the competitive results with the previous state-of-the-art models on the two datasets under all metrics, which verify the effectiveness of our approach.

\smallskip\noindent\textbf{Human Evaluation} \ 
Moreover, we invite two professional radiologists to evaluate the perceptual quality of 200 randomly selected reports generated by our AlignTransformer and the R2Gen \cite{Chen2020Generating}.
The radiologists are unaware of which model generates these reports and are encouraged to select a more accurate report for each pair.
The results of last column in Table~\ref{tab:main_result} show that our approach is better than R2Gen in clinical practice with winning pick-up percentages.

Overall, the results verify that our proposed framework can generate accurate and proper reports, which could assist radiologists in making decisions.

\begin{table}[t]
\centering
\scriptsize
\caption{Ablation study of our method, which includes the AHA and the MGT. The $N$ denotes the number of alignments in AHA. The `Baseline' and `w/ AHA+MGT' denote the ResNet-50 image encoder \cite{he2016deep} equipped with the Transformer report decoder in \cite{Chen2020Generating} and our proposed AlignTransformer, respectively.
\label{tab:ablation}}
\setlength{\tabcolsep}{2pt}
\begin{tabular}{@{}l l c c c c c c c@{}}
\toprule

Datasets & Methods & $N$ & BLEU-1 & BLEU-2 & BLEU-3 & BLEU-4 & METEOR & ROUGE-L\\
\midrule [\heavyrulewidth]
\multirow{6}{*}[-3pt]{IU-Xray} 
& Baseline & - & 0.442 & 0.285 & 0.195 & 0.138 & 0.176 & 0.357 \\
\cmidrule(){2-9}
& w/ AHA & 1 & 0.467 & 0.301 & 0.210 & 0.159 & 0.185 & 0.368 \\
& w/ AHA & 2 & 0.473 & 0.305 & 0.217 & 0.163 & 0.188 & 0.372 \\
& w/ AHA & 3 & 0.476 & 0.308 & 0.219 & 0.164 & 0.192 & 0.374 \\
& w/ AHA & 4 & 0.471 & 0.303 & 0.215 & 0.160 & 0.187 & 0.370\\
\cmidrule(){2-9}
& w/ AHA+MGT  & 3 & \bf 0.484 & \bf 0.313 & \bf 0.225 & \bf 0.173 & \bf 0.204 & \bf 0.379\\ 
\bottomrule
\end{tabular}
\end{table}

\section{Analysis}
In this section, we conduct analysis on the benchmark IU-Xray dataset \cite{Dina2016IU-Xray} from different perspectives to better understand our proposed framework.

\smallskip\noindent\textbf{Ablation Study}
We investigate the contribution of each component in our approach.
Table~\ref{tab:ablation} shows that our AHA can significantly promote the performance over all metrics, with up to 18.8\% gain in BLEU-4 ($N=3$).
This indicates that the disease-grounded visual features extracted by our AHA contain sufficient accurate abnormal information, which is vital in alleviating the data deviation problem and improving the performance of medical report generation.
For the MGT, Table~\ref{tab:ablation} shows that the MGT can further boost the performance under all metrics, which demonstrates the effectiveness of our MGT to adaptively exploit the multi-grained disease-grounded visual features.

\smallskip\noindent\textbf{Visualization and Example}
In Fig.~\ref{fig:introduction}, we give the visualization of the AHA and an intuitive example to better understand our approach.
As we can see, for the current state-of-the-art model R2Gen \cite{Chen2020Generating}, it produces some inexact descriptions and repeated sentences of normalities, which is due to the overwhelming normal images in the dataset, i.e., data deviation \cite{Shin2016Learning,Jing2019Show,fenglin2021PPKED,fenglin2021CMCL}.
Since our AHA can efficiently capture the abnormal regions by aligning visual regions and disease tags, we can generate more accurate report for the abnormal regions. For example, the generated report correctly describes ``\textit{The lungs are hyperlucent}'', ``\textit{There is left lower lobe airspace disease suspicious for pneumonia}'' and ``\textit{There is a moderate left sided pleural effusion}''.

\section{Conclusions}
In this paper, we propose an AlignTransformer to alleviate the data bias problem and model the very long sequence for medical report generation. The experiments prove the effectiveness of our method, which not only generates meaningful medical reports with accurate abnormal descriptions and regions, but also achieves competitive results with state-of-the-art models on two public datasets with the best human preference. The results justify the usefulness of our framework in assisting radiologists in clinical decision-making, reducing their workload.

\section*{Acknowledgments}
We sincerely thank all the anonymous reviewers and chairs for their constructive comments and suggestions that substantially improved this paper.
Xian Wu is the corresponding author of this paper.

%
%
%
\bibliographystyle{splncs04}
\bibliography{miccai2021}

\end{document}